\documentclass[12pt]{article}
\usepackage{latexsym,graphicx}
\usepackage{subfigure}
\usepackage{amssymb}
\usepackage{amsmath}
\usepackage{amscd}
\usepackage{amsthm}
\usepackage{float}
\usepackage[left=2cm,top=2.5cm,right=2.5cm,bottom=1.5cm]{geometry}
\usepackage{times}
\theoremstyle{definition}

\usepackage{xcolor}
\begin{document}
\begin{center}
\large{\bf{ An oscillating Rastall universe crossing the phantom divide line}} \\
\vspace{10mm}
\normalsize{ Nasr Ahmed $^{1,2}$, Anirudh Pradhan$^{3}$, Archana Dixit $^{4}$ }\\
\vspace{5mm}
\small{\footnotesize $^1$ Mathematics Department, Faculty of Science, Taibah University, Saudi Arabia.} \\
\small{\footnotesize $^2$ Astronomy Department, National Research Institute of Astronomy and Geophysics, Helwan, Cairo, 
	Egypt}\\
\small{\footnotesize $^3$ Centre for Cosmology, Astrophysics and Space Science (CCASS), GLA University, Mathura-281 406, Uttar Pradesh, India}\\
\small{\footnotesize $^4$  Department of Mathematics, Gurugram University, Gurugram-122003, Haryana, India}\\

\vspace{2mm}
$^{1,2}$E-mail: nasr.ahmed@nriag.sci.eg \\
$^3$E-mail:  pradhan.anirudh@gmail.com \\
$^4$E-mail: archana.ibs.maths@gmail.com
\end{center}   
\begin{abstract} 
A cyclic flat universe with quintom behavior and future big rip has been presented in the framework of Rastall gravity, which is an extension of the standard $\Lambda$CDM model. The Hubble parameter oscillates periodically between positive and negative values from one cycle to the next. Cosmic transit has been simulated through an oscillating time-dependent deceleration parameter, and is expected to occur at approximately $ 8.7~~ \text{Gyr}$. The causality is satisfied all the time except near the initial singularity and the future Big Rip singularity.The apparent horizon, entropy and other thermodynamical quantities associated to the current model have been analyzed. Energy conditions have been investigated.  
\end{abstract}
\section{Introduction and Motivation} \label{secintro}
One of the most significant and fascinating phenomena in contemporary cosmology is the universe's accelerating expansion. Our understanding of the composition and development of the cosmos underwent a paradigm change when it was discovered in the late 1990s through the observation of Type Ia Supernovae \cite{1a,1b}. Numerous independent observations, including as large-scale structural surveys and measurements of cosmic microwave background radiation, have supported this ground-breaking discovery \cite{1c,1d,1e}.
At both large scales and strong curvature, Einstein theory of General Relativity is inconsistent with the observations. Cosmic acceleration has been a basic motivation behind modified gravity theories since it can't be explained by GR without assuming dark energy \cite{1b,1f}. Examples of such modified gravity theories are $f(R)$ gravity \cite{noj1}, Gauss-Bonnet gravity
 \cite{noj8,noj81} where the Gauss-Bonnet term $\mathcal{G}=R^2-4R^{\mu\nu}R_{\mu\nu}+R^{\mu\nu\rho\delta}R_{\mu\nu\rho\delta}$ has been used in the action instead of the Ricci scalar $R$, and f(T) gravity \cite{torsion} where the torsion scalar ($T$) has been used. the generalized $f(R,T)$ gravity where $T$ is the trace of the energy-momentum tensor. More examples include $\kappa(R,T)$ gravity \cite{29a} and extra-dimensional gravity theories \cite{30,31,32,hol}. \par

Bouncing scenarios (also named as cyclic cosmology) have been firstly suggested as an alternative way to solve some of the problems of the standard model of cosmology such as flatness and initial singularity. The universe in the singularity-free Big Bounce emerges from a previous contracting stage \cite{bounc2,bounc3,bounc4,bounc5} (see \cite{bounc6} for a review) where the contraction-expansion cycle is supposed to continue forever. There have been an extensive interest in cyclic cosmological models in the literature. they have been studied in many modified gravity theories \cite{bounc4,bounc5,bounc7,bounc7a,bounc7b,bounc8,bounc9,nasrnt,asa}. A general cyclic model in $f(R)$ gravity has been constructed in \cite{33a}. A unitary version of Conformal Cyclic Cosmology ( CCC ) has been suggested in \cite{33g}. An M-theory inspired cyclic Model with two branes has been presented in \cite{58} where the EoS parameter satisfies $\omega \gg 1$ during the contraction phase. A new novel cyclic theory of the universe has been presented in \cite{new} where the scale factor grows exponentially from one cycle to another. This recent cyclic model resolves many problems and produces a nearly scale invariant spectrum of density perturbations.
An interesting singularity-free oscillatory model has been discussed in \cite{pr1} where the late-time evolution was probed in the context of Quasi Steady State cosmology QSSC \cite{20,21,22} to alleviate the Hubble tension. The cosmic scale factor parametrization has been chosen according to the description in \cite{20,21,22,23,24} as  
\begin{equation} 
a(t)=e^{t/P} \left[1+\eta \cos (2\pi t/Q)\right].
\end{equation} 
Here $\eta$ is a dimensionless constant while $P$ and $Q$ have dimensions of time. In the limit when the dimensionless constant $\eta$ goes to zero, A de Sitter-like evolution occurs. When time goes to zero, the scale factor tends to the value $1+\eta$ where singularity doesn't exist.\par 

Rastall gravity is one such modification that was proposed by Peter Rastall in 1972 \cite{A1}. Due to its presentation as a modified theory of gravity with a non-conserved stress energy tensor and a peculiar non-minimal coupling between geometry and matter, Rastall gravity (RG) has emerged as an interesting and important theory of gravity in modern times. 
The conservation law that demonstrated a divergence free energy momentum tensor in general relativity (GR) has been accepted; that is, $\nabla_{\nu}T^{\mu\nu} = 0$, where $\nabla_{\nu}$ is the covariant derivative. Rastall, on the other hand, adopted a new and distinct conservation law that conjectured the GR concepts.
In this theory, the energy momentum tensor does not a conserved
 quantity. Rastall formulation is defined as $\nabla_{\mu} T^{\mu}_{\nu}= \left(\frac{\kappa}{16 \pi}\right)\nabla_{\nu} R$, where $\kappa$  is the
coupling constant and $R$ be the Ricci scalar. The second Bianchi identity in
this theory remains the same i.e., $\nabla^{\mu} G_{\nu\mu} = 0$, where $G_{\mu\nu} = R_{\mu\nu} - \frac{1}{2} R g_{\mu\nu}$ be
the Einstein tensor. One significant aspect of RG is that, while the mathematical portion of the theory remains invariant, modifications have only been made to the matter portion \cite{A2,A3,A4}. For the Rastall gravity framework, Visser \cite{A4a} have claimed that the Rastall gravity may have the complete equivalence with the General relativity. However, Darabi et al \cite{A4b} have discussed that the Rastall gravity may be visualized as a modified gravity theory and it is different from the General relativity. Das  \cite{A5} demonstrated that RG is equal to Einstein gravity and offered some cosmological implications in the structure of modified RG. Recently, Golovnev \cite{A5a} have claimed that Rastall gravity may be equivalent to the General relativity by analyzing the claims of Visser \cite{A4a} as well as Darabi et al \cite{A4b}. The derivation of action for Rastall gravity have been given subjected to the condition $\sqrt{-g}=1$ \cite{A5a}. Holographic dark energy in RG was introduced by Ghaffari \textit{et al.} \cite{A6}, taking into account vacuum energy, which serves as DE. They accepted the idea that the current accelerating cosmos is caused by the sum of this energy and the Rastall term. With the assumption that the universe is made up of interacting/non-interacting dark energy (DE) and dark matter (DM), Saleem and Shahnila \cite{A7} investigated the phenomenon of cosmic evolution using curved FLRW space-time bounded by an apparent horizon with a particular holographic cut-off in the framework of Rastall gravity.
Saleem \textit{et al.} \cite{A8} recently introduced constant-roll warm inflation, a novel method for determining the precise inflationary solutions to the Friedman equations within the framework of Rastall theory of gravity (RTG). Using a linear parametrization of the Equation of State (EoS) in FLRW background, Singh \textit{et al.} \cite{A9} examined the evolution of the universe within the framework of RTG. In the framework of RTG, Saleem and Hassan \cite{A10} studied the dynamics of warm inflation induced by vector fields and concluded that the modified theory was compatible with the 2018 Planck observational data. In various contexts, several researchers have examined the cosmological scenario within the framework of Rastall gravity \cite{A11,A12,A13,A14,A15,A16,A17,A18,A19a,A19b,A19}. This theory have received a widespread attention to discuss and explain different phenomenon related to the universe evolution. As a fact, we aim to study the universe evolution exhibiting oscillatory expansion with analysis of thermodynamic behavior. It is worthwhile to mention that Rastall gravity possess oscillatory solutions subjected to the fixed point which are marginally stable \cite{A17,A18}.
\par

Models of dark energy that permit the equation of state parameter, w, to cross the ``phantom divide line" (w=-1) investigate possibilities in which the expansion of the universe can go from accelerating to accelerating at an ever-increasing rate, which could result in a ``Big Rip" \cite{R1,R2}.  In Weyl-type f(Q,T) gravity, the authors \cite{R2} analyze the behavior of the effective equation of state (EoS) parameter and find that the best fit values of the model parameters support a crossing of the phantom divide line $(\omega_{eff} = -1)$ to go from $(\omega_{eff} > -1)$ (quintessence phase) to $(\omega_{eff} =<-1)$ (phantom phase). The long-term destiny of the universe depends on our ability to comprehend dark energy and its behavior, especially the potential for crossing the phantom divided line. \cite{R3} discusses the conditions required to cross the phantom divide line in a closed Friedmann-Robertson-Walker universe using an interacting holographic dark energy model. The authors \cite{R4} reconstitute the dark energy parameters from the most recent 397 Sne Ia, CMB, and BAO using three distinct parameterized dark energy models with the specified current matter density, $\Omega_{m0}$.  They discover that an evolving dark energy with a crossing of the phantom dividing line is preferred when $\Omega_{m0}$ is not modest, such as when $\Omega_{m0}= 0.28 $ or $0.32$.To help the reader understand, we quote a few helpful papers \cite{R5,R6,R7,R8,R9,R10}. 
\par 
The acceleration of the expansion of the universe, which defies the predictions of General Relativity (GR) without including dark energy, is the driving force behind this investigation.  As a result, modified gravity theories have been examined, such as Rastall gravity, which adds a non-minimal link between geometry and matter and a non-conserved stress-energy tensor.  The goal of the authors' cyclic universe model is to address problems such as the initial singularity and future as big rip scenarios, in addition to quintom behavior, which is the transition between quintessence and phantom phases with a crossing of the phantom dividing line. 
\par
An oscillating Hubble parameter that alternates between positive and negative values across cycles is the result of the authors' proposed periodic parametrization of the deceleration parameter in the current model.  With an anticipated occurrence at roughly 8.7 Gyr, this periodic behavior effectively simulates cosmic transit by suggesting that the cosmos experiences periods of acceleration and deceleration.  Aside from the area around the initial singularity and the future big rip, the model also satisfies causality. To make sure the model is physically feasible, thermodynamical characteristics such as entropy and the energy conditions are examined. 
\par 

Due to the ad hoc periodic parametrization of the deceleration parameter $q$ with the correct sign flipping as $ q=m \cos{kt} -1$ \cite{2},  the universe accelerates after an epoch of deceleration (for every single cycle) which agrees with recent observations. Such periodic form of $q$ leads to the following forms for the Hubble parameter and scale factor 
\begin{equation} \label{qha}
 H=\frac{k}{m \sin{kt} +c}~~,~~a=a_0\left[\tan(\frac{1}{2}kt)\right]^{\frac{1}{m}},
\end{equation} 
where $m>0$, $k>0$ and $k$ acts as a cosmic frequency parameter. This specific form of the deceleration parameter has firstly been introduced in \cite{2} where a new oscillating Quintom Model has been constructed. Using the well known redshift relation $z=\frac{1}{a}-1$ we get
\begin{equation}
t=\frac{2}{k} \tan^{-1}{\frac{1}{a_0^m(z+1)^m}}.
\end{equation}
Observations suggests that the signature change of $q$ occurs at $z=0.64$ for $m=1.55$ \cite{3,4,5,6}. Since cosmic transit happens when $q=0 ~(i.e. ~\ddot{a}=0)$, we have
\begin{equation}
t_{q=0}=\frac{1}{k} \cos^{-1}{\frac{1}{m}} \approx 8.7~~ \text{Gyr} ~~~~~~~~~~~~for~~ m=1.55 ,~~ k=0.1
\end{equation}
The scale factor (\ref{qha}) suffers from future Big Rip singularity where the slope of $a(t)$ increases hugely and goes to infinity $\frac{da}{dt}\rightarrow \infty \Rightarrow$ which eventually tearing apart the space-time fabric .

\section{Framework of the Model }
Our goal is to obtain solutions by carefully taking into consideration the matter source and the gravitational background.
The basic idea behind Rastall gravity \cite{A1} is that the energy-momentum conservation in GR can't be always valid in curved space-time  and, instead of $T^{\mu\nu}_{;\mu}=0$, we should have  
\begin{equation}
T^{\mu\nu}_{;\mu}=\lambda R^{,\nu},
\end{equation}
which leads to the generalized field equations
\begin{equation} \label{generalized}
G_{\mu\nu}+K\lambda g_{\mu\nu}R=KT_{\mu\nu}.
\end{equation}
GR is recovered for $\lambda=0$. $G_{\mu\nu}$ is the Einstein tensor and $K$ is a coupling constant. The FRW metric given by
\begin{equation}
ds^{2}=-dt^{2}+a^{2}(t)\left[ \frac{dr^{2}}{1-\kappa r^2}+r^2d\theta^2+r^2\sin^2\theta d\phi^2 \right] ,\label{RW}
\end{equation} 
where $a(t)$ is the scale factor and $\kappa$ equals $0$ for a flat universe. Applying equation (\ref{generalized}) to the metric (\ref{RW}) we obtain the cosmological equations as
\begin{equation} \label{1}
3(1-4K\lambda)H^2-6K\lambda \dot{H}+3(1-2K\lambda)\frac{\kappa}{a^2}=K\rho,
\end{equation}
\begin{equation}
3(1-4K\lambda)H^2+2(1-3K\lambda) \dot{H}+(1-6K\lambda)\frac{\kappa}{a^2}=-Kp.
\end{equation}
In the current work, we will be interested only in the observationally supported flat case where $\kappa=0$ \cite{teg, ben,sp,nasrent}. The energy density $\rho$, cosmic pressure $p$ and EoS parameter $\omega$ are written as
\begin{eqnarray}
\rho&=&-\frac{1}{K}\left(12\lambda K H^2 +6\lambda K\dot{H}-3H^2\right),\label{9}\\
p&=&\frac{1}{K}\left(12\lambda K H^2 +6\lambda K\dot{H}-3H^2 -2\dot{H}\right), \label{10}
\end{eqnarray} 
where the dot denotes differentiation w.r.t time and the EoS parameter $\omega=p/\rho$. Equations (\ref{9}) and (\ref{10}) show that
\begin{equation}
p=-\rho-\frac{2}{K} \dot{H},
\end{equation}
where $\dot{H}$ expresses the rate of change of the Hubble parameter as
\begin{equation}
\dot{H}=-\frac{K}{2}(\rho+p)=-\frac{K}{2}(\gamma \rho).
\end{equation}
We have considered the EoS $p=(\gamma-1)\rho$ with $\frac{2}{3} \leq \gamma \leq 2$. As has been pointed out in \cite{rastall1}, inserting this equation for $\dot{H}$ in (\ref{1}) for the flat case results in .
\begin{equation}
H^2=\frac{K\rho(3K\lambda \gamma -1)}{3(4k\lambda-1)}.
\end{equation}

The physical behavior of energy density has been plotted in figure (1). The pressure changes sign from positive in a decelerating era to negative in an accelerating era dominated by dark energy with negative pressure. As a function of redshift, The EoS parameter $\omega(z)$ meets $-1$ at $z=0$. The time evolution of the EoS parameter verses reveals a Quintum behavior as it passes the phantom divide line at far future. The Hubble parameter $H$ is decreasing during the expansion as $\dot{H} < 0$, while it is increasing during the contraction where $\dot{H}>0$. The general dynamical behavior can be deduced from the negative values of $q$. In each cycle, the EoS parameter $\omega(t)$ lies between $-2.25$ and $\frac{1}{3}$ which agrees with observations \cite{ob1,ob2}. It starts from a positive value (radiation-like era), keeps decreasing to zero (dust era $\omega = 0$) and passes to the negative values. After reaching the DE-dominated era, it crosses the cosmological constant boundary to the phantom era where $\omega<-1$ possessing a Quintom feature. The Quintom dynamics associated with the crossing of the phantom divide line leads to $\omega<-1$ today is also observationally supported \cite{quintom}. A cosmological bouncing Quintom model has been studied in \cite{quintom2}. $\omega$ is an increasing function in the second half of the cosmic cycle starting from $-2.25$. Some observations favor DE with $\omega$ less than $-1$ \cite{vikman,obb1, obb2}. The idea whether DE can evolve to the phantom era or not has been extensively investigated. In \cite{vikman} it has been shown that, in a DE dominated universe, the transition of DE from $\omega \geq -1$  to $\omega <-1$ can't be explained by models of classical scalar fields dynamics unless more complicated physics is included \cite{d1,d2,d3}.
\section{Work Density and Entropy}
In this section, we calculate and analyze some thermodynamical quantities. The $(n+1)$-dimensional FRW metric is written as
\begin{equation}
ds^{2}=-dt^{2}+a^{2}(t)\left( \frac{dr^{2}}{1-\kappa r^2}+r^2d \Omega^2_{n-1} \right), \label{nRW}
\end{equation} 
where the line element of an $(n-1)$-dimensional unit sphere is denoted by $\Omega^2_{n-1}$. This metric can also be written as \cite{Bak}
\begin{equation}
ds^{2}=h_{ab}dx^adx^b+\widetilde{r}^2\Omega^2_{n-1},
\end{equation}
where $\widetilde{r}=a(t)r$, $x^0=t$, $x^1=r$ and the 2-dimensional metric $h_{ab}=diag(-1,\frac{a^{2}}{1-\kappa r^2})$. The apparent horizon is defined geometrically as an imaginary surface beyond which null geodesic congruences recede from the observer \cite{mainref}. It is a dynamical structure which evolves with time and is determined by the relation $h_{ab} \partial_a \widetilde{r} \partial_b \widetilde{r}=0$. This leads to the expression of the apparent horizon's radius in terms of the Hubble parameter as
\begin{equation}
\widetilde{r}_A=\frac{1}{\sqrt{H^2+\frac{\kappa}{a^2}}}.
\end{equation}
For a flat universe (the case we are considering here), The apparent horizon's radius is the inverse of the Hubble parameter $1/H$ which is the same definition of the Hubble horizon $\widetilde{r}_H$. That means 
\begin{equation}
\widetilde{r}_A=\widetilde{r}_H=\frac{1}{H},   ~~~ \text{for}~~~ \kappa=0.
\end{equation}
Consequently, the rate of change of the horizon radius $\dot{\widetilde{r}_A}=-\frac{\dot{H}}{H^2}=-\widetilde{r}_A^2\dot{H}$ or can be written in general as 
\begin{equation}
\dot{\widetilde{r}_A}=-\widetilde{r}_A^3 H  \left(\dot{H}-\frac{\kappa}{a^2}\right).
\end{equation}
\begin{figure}[H]
  \centering            
	\subfigure[$q(z)$]{\label{F482}\includegraphics[width=0.3\textwidth]{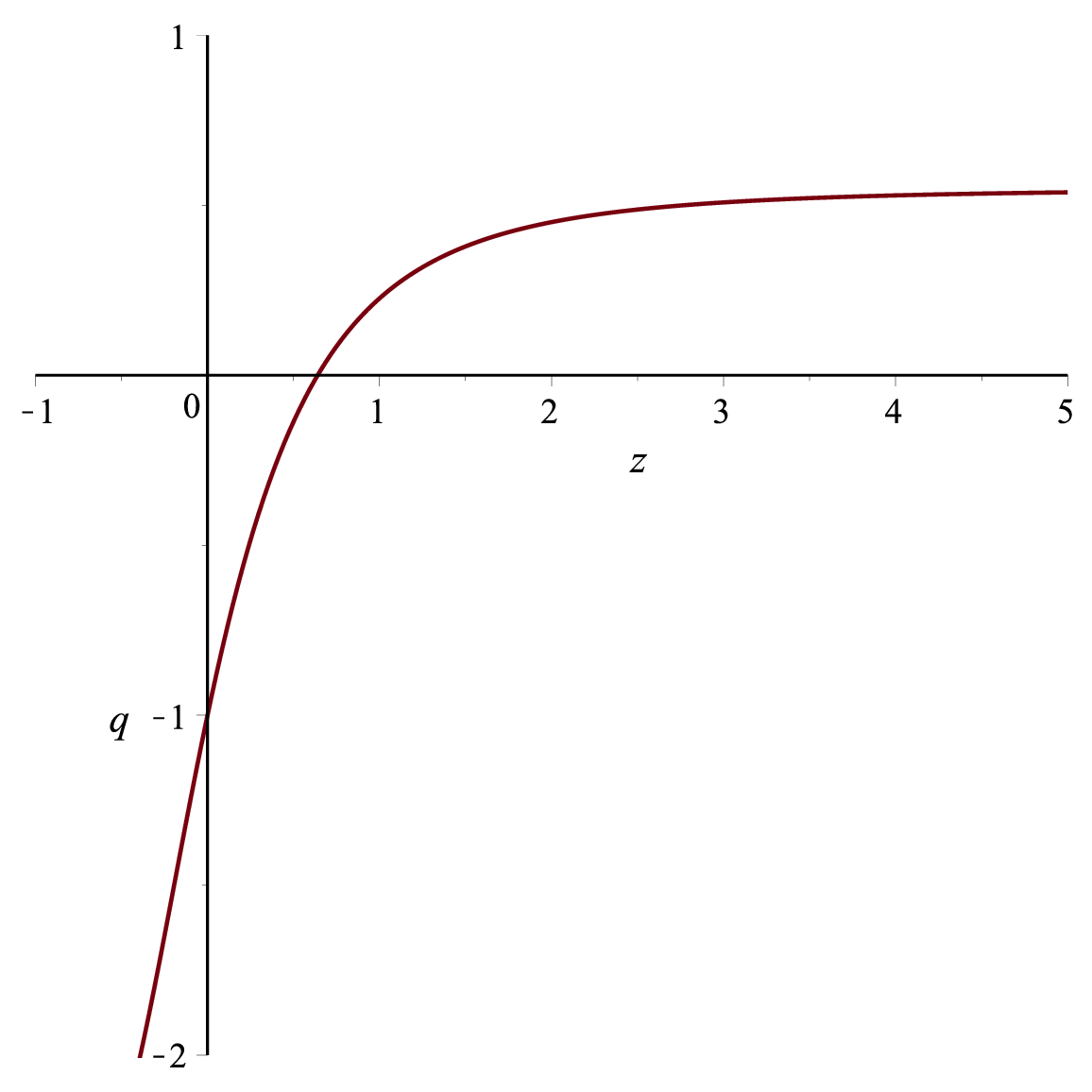}}\hspace{5mm}
		\subfigure[$H(z)$]{\label{F48}\includegraphics[width=0.3\textwidth]{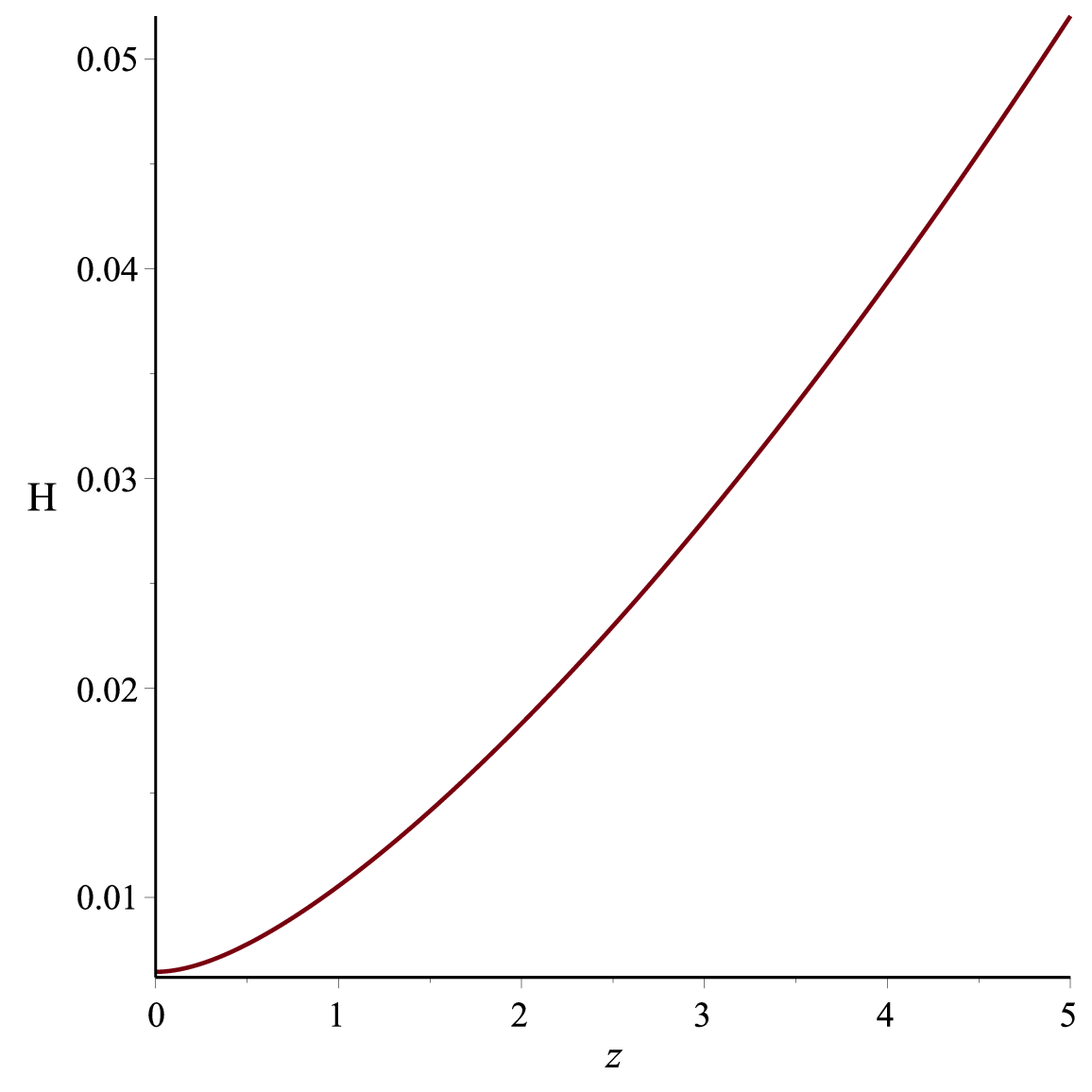}}\hspace{5mm}
	 \subfigure[$\rho(z)$]{\label{F06}\includegraphics[width=0.3\textwidth]{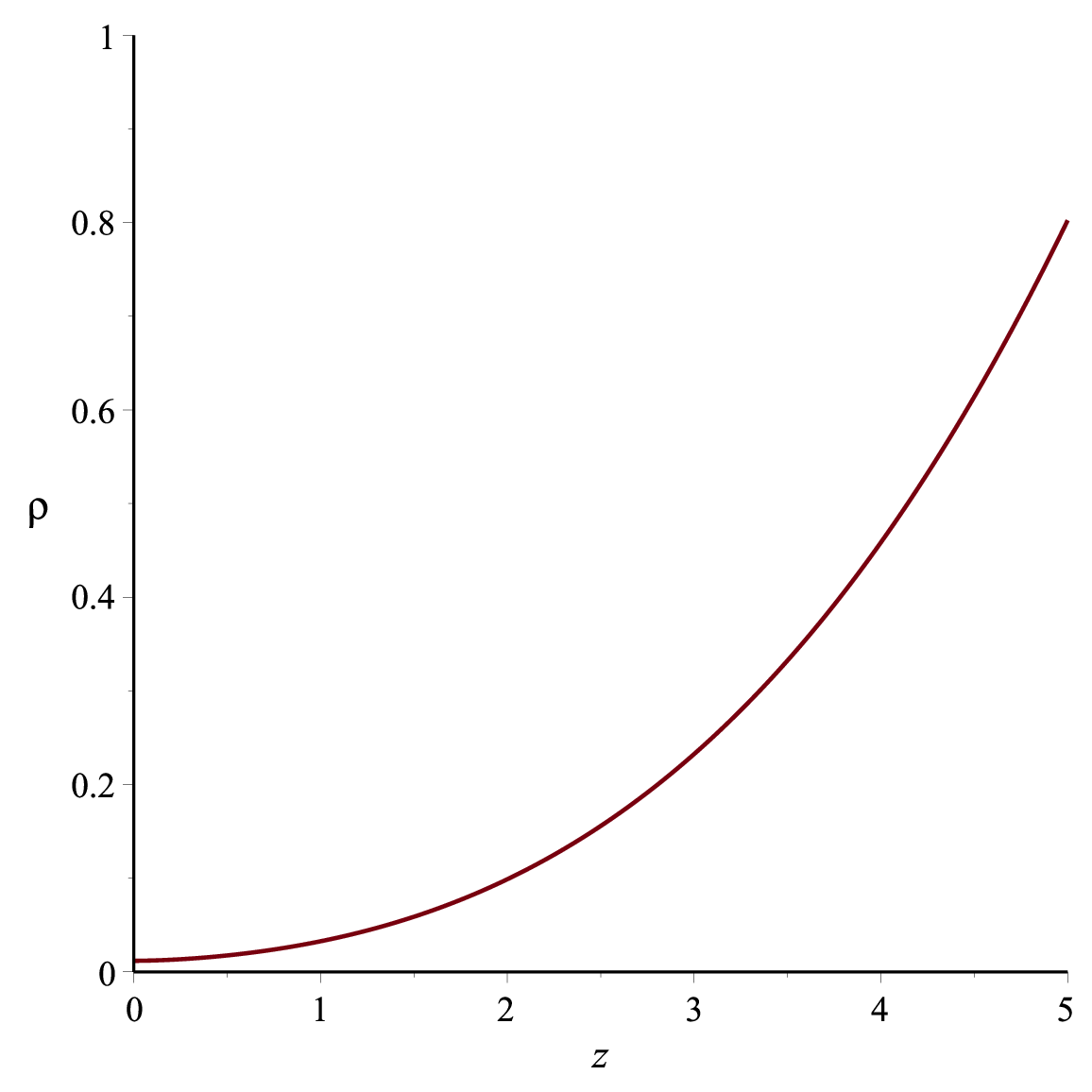}}\\
	 \subfigure[$p(z)$]{\label{F67}\includegraphics[width=0.3\textwidth]{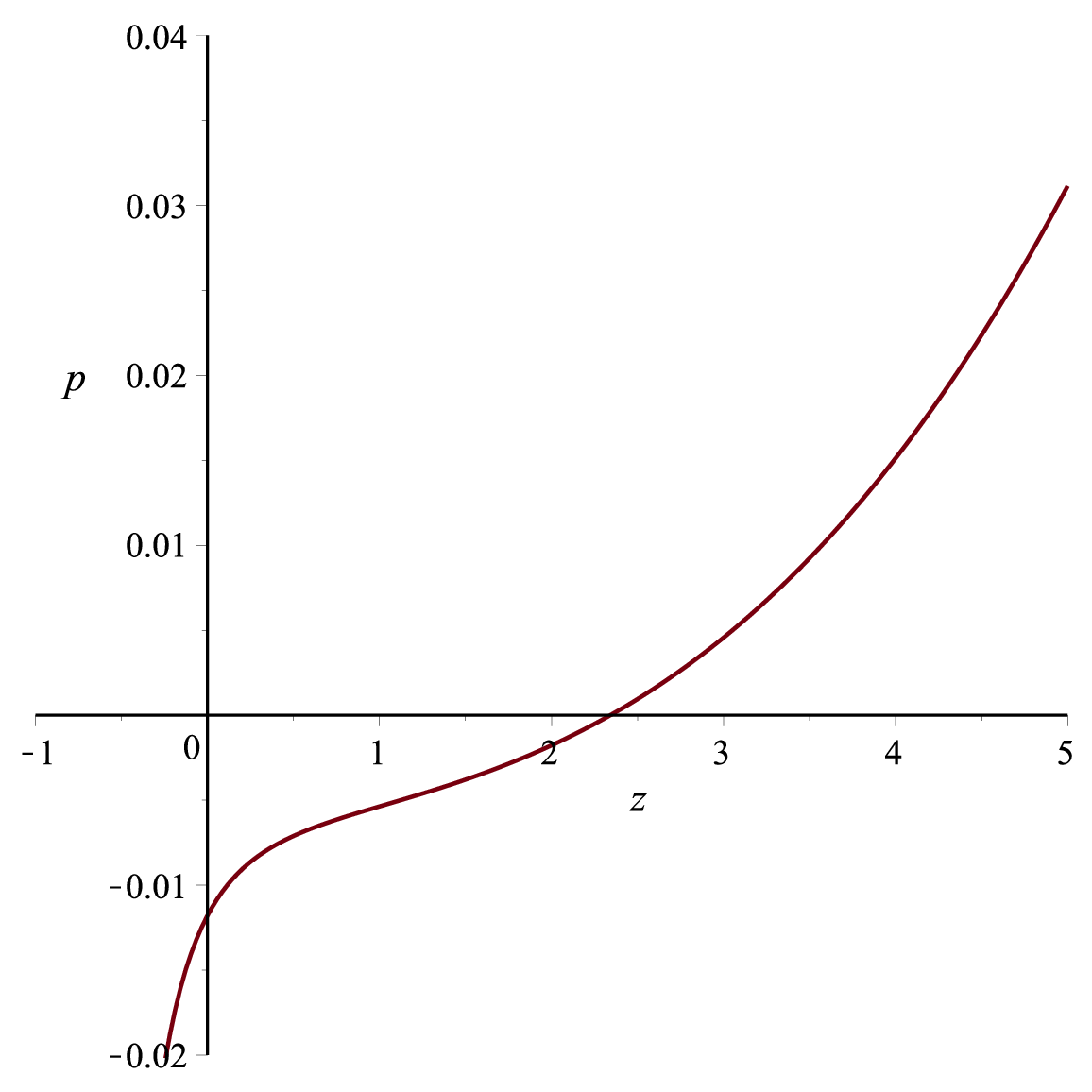}} 	\hspace{5mm}
		 \subfigure[$\omega(z)$]{\label{F8}\includegraphics[width=0.3\textwidth]{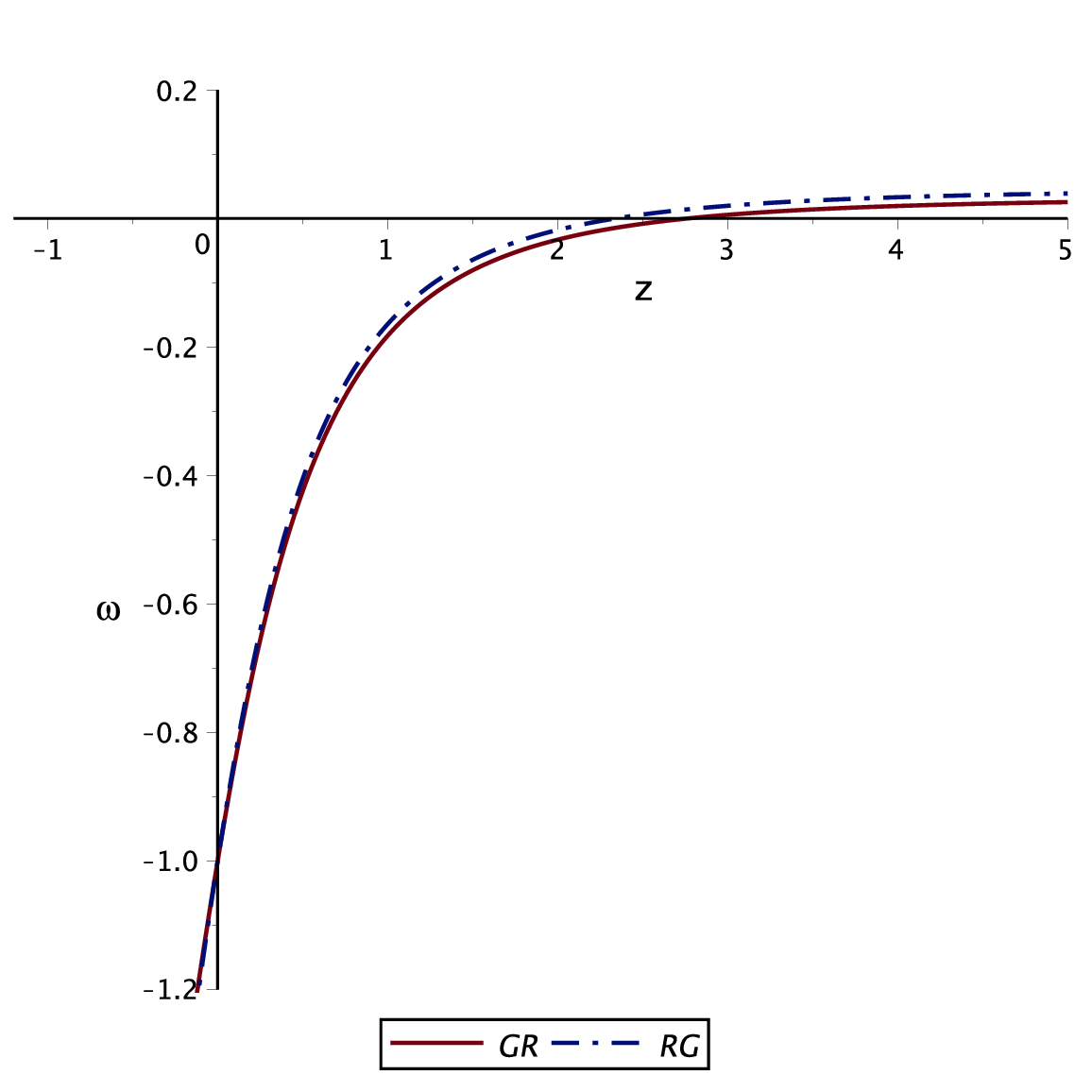}} 	\hspace{5mm}
		 \subfigure[$\omega(t)$]{\label{Fb}\includegraphics[width=0.3\textwidth]{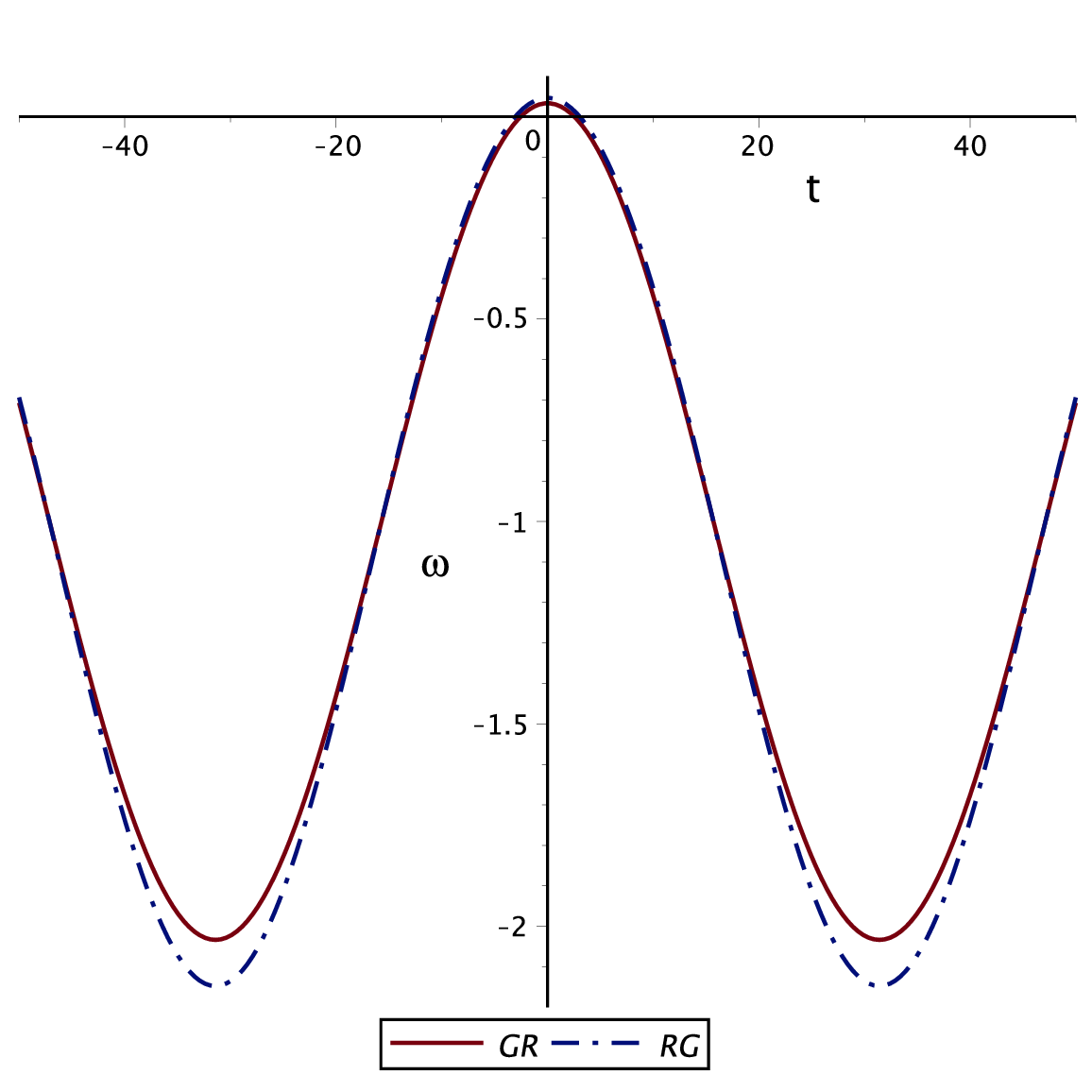}}\\
			\subfigure[$\dot{H}$]{\label{F4}\includegraphics[width=0.3\textwidth]{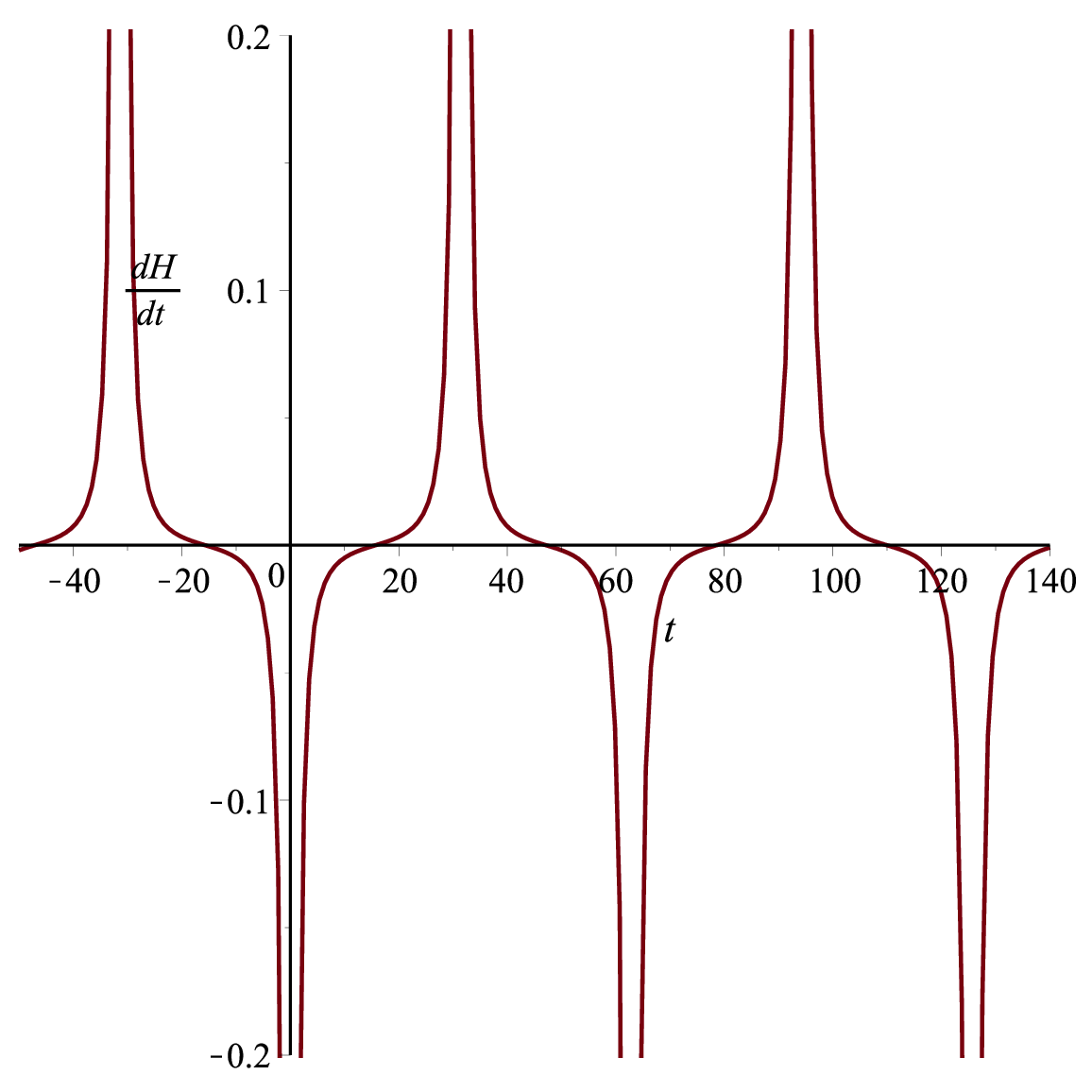}}	\hspace{5mm}
		\subfigure[$H(t)$]{\label{F82}\includegraphics[width=0.3\textwidth]{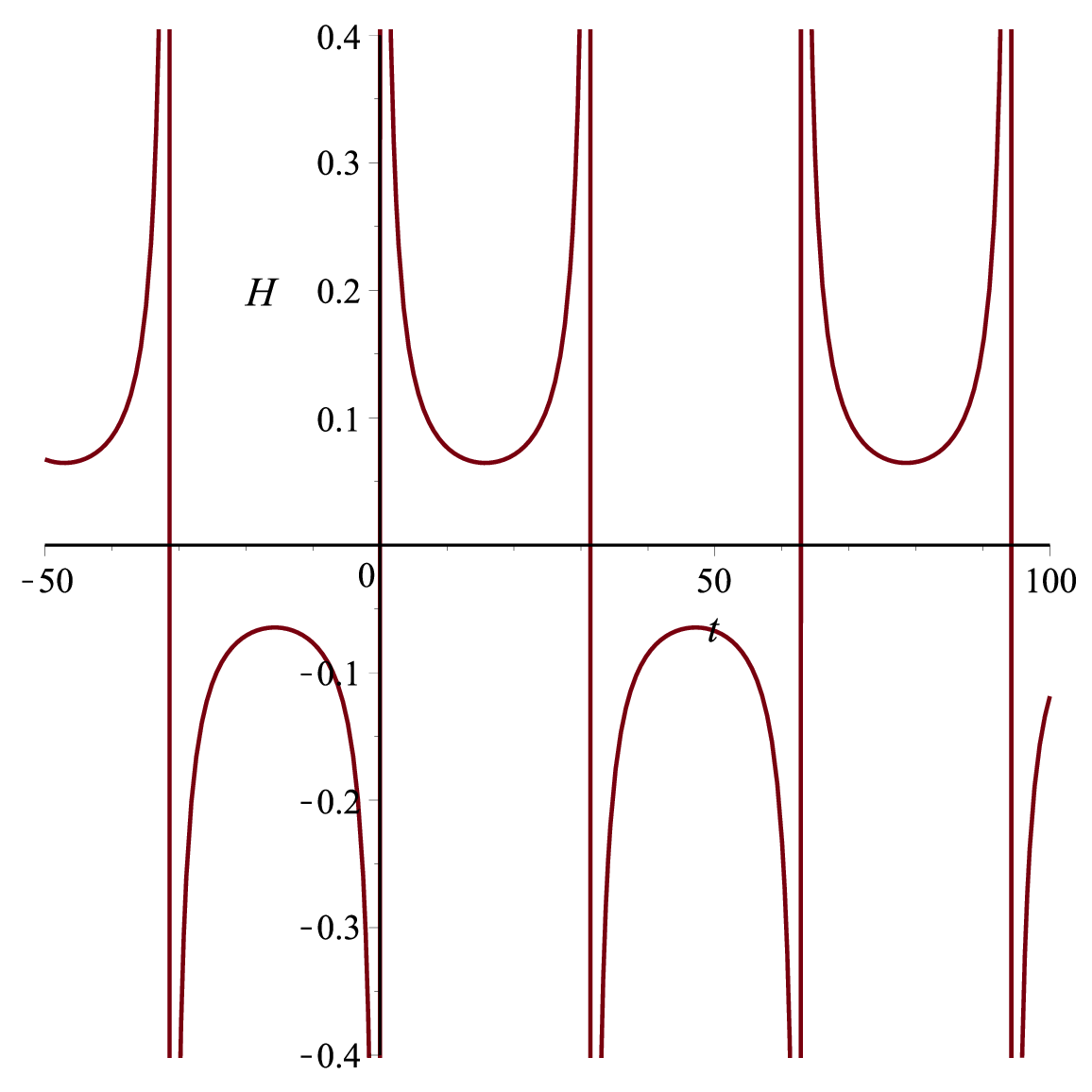}}\hspace{5mm}
		\subfigure[$a(t)$]{\label{F8t2}\includegraphics[width=0.3\textwidth]{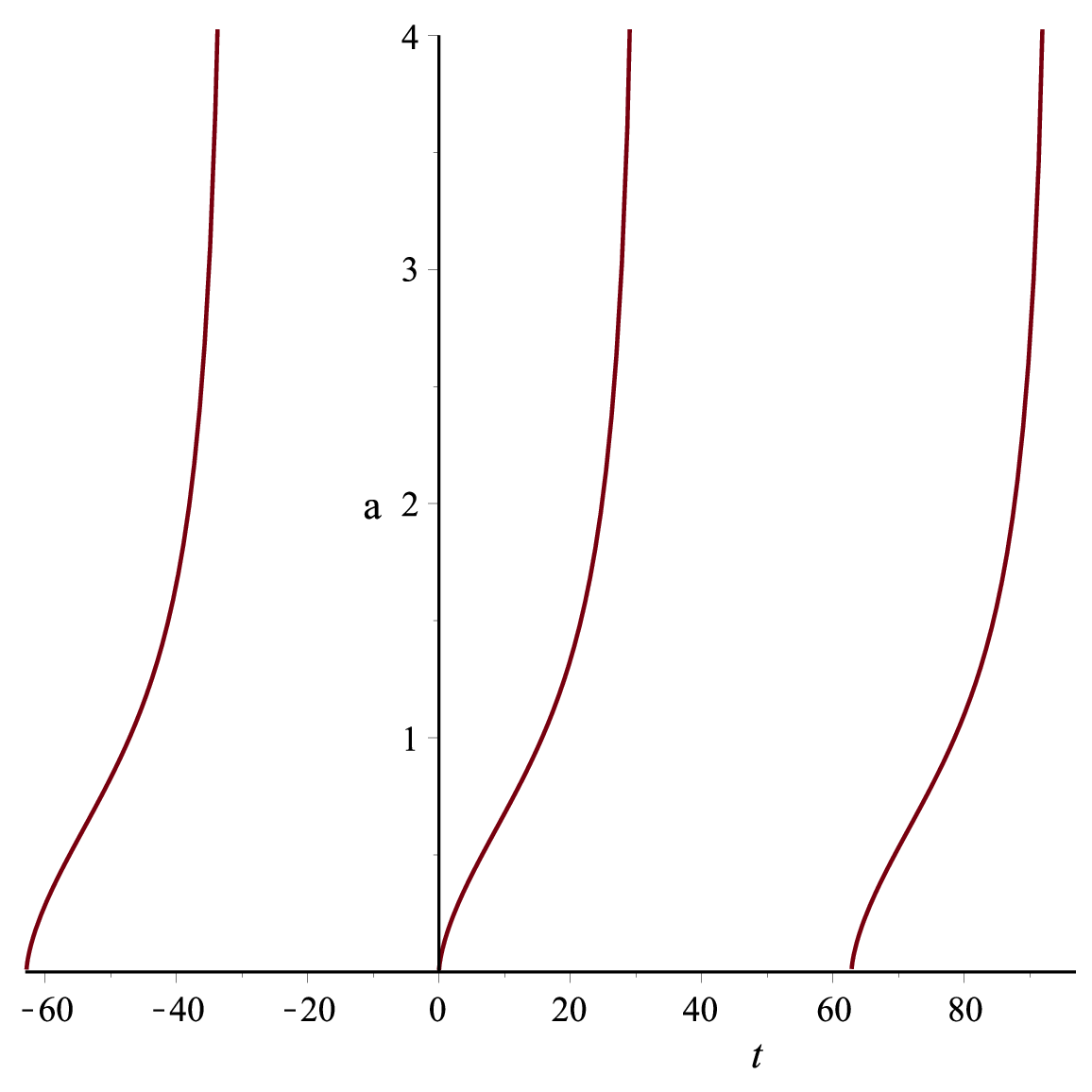}}
		 \caption{ Evolution of the model parameters against the redshift $z$ (a) The DP $q(z)$ changes sign for $m>1$ and equals $-1$ at $z=0$. (b) The Hubble parameter $H(z)$. (c) The physical behaviour of energy density $\rho(z)$. (d) A sign flipping in the evolution of cosmic pressure $p(z)$ (e) The EoS parameter $\omega(z)$ equals to $-1$ at $z=0$. (f) $\omega(t)$ evolution shows a Quintum behaviour where it crosses the phantom divide line at very late times. GR stands for General Relativity and RG for Rastall Gravity. (g) The rate of change of the Hubble parameter $\dot{H} < 0$ during the expansion which means a decreasing $H$, while this rate is $> 0 $ during the contraction. (h) The Hubble parameter $H > 0$ during the expansion, and $H < 0$ during the contraction. (i) The evolution of scale factor shows an enormous increase over a short period of time (a future big rip ). The unit of time $t$ is taken in Gyr. Here $m=1.55$, $\lambda=1.4$, $K=0.01$, $k=0.1$ and $a_0=1$. }
  \label{fr}
\end{figure}

The work density $W$ is defined as the work done by the cosmic volume change due to the change of the radius of the apparent horizon. For the current model we get
\begin{eqnarray}
W&=&-\frac{1}{2} T^{ab}h_{ab}=\frac{1}{2}(\rho-p)\\
&=&-\frac{1}{K}  \left(12\lambda K H^2 +6\lambda K\dot{H}-3H^2-\dot{H}\right).
\end{eqnarray}
In the case of the FRW spacetime with a perfect fluid. The surface gravity $\kappa_{sg}$ is also defined as 
\begin{equation}
\kappa_{sg}=\frac{1}{2\sqrt{-h}}\partial_a(\sqrt{-h}h^{ab}\partial_b \widetilde{r}_A)=-\frac{1}{\widetilde{r}_A}\left( 1-\frac{1}{2}\frac{\dot{\widetilde{r}_A}}{H\widetilde{r}_A} \right),
\end{equation}
where $h$ is the determinant of $h_{ab}$. In terms of $\kappa_{sg}$, the temperature on the horizon is given as $\frac{|\kappa_{sg}|}{2\pi}$
\begin{equation}
T_A=\frac{1}{2\pi \widetilde{r}_A}\left| 1-\frac{1}{2}\frac{\dot{\widetilde{r}_A}}{H\widetilde{r}_A} \right|.
\end{equation}
So, the temperature depends on the Hubble parameter $H$, The radius of the apparent horizon $\widetilde{r}_A$ and its rate of change during evolution $\dot{\widetilde{r}_A}$ which is a result of the dynamical nature of the apparent horizon. If the change in the apparent horizon is very slow, then $\frac{\dot{\widetilde{r}_A}}{H\widetilde{r}_A} \ll 1$ and we obtain the expression for Hawking temperature $T_H=\frac{1}{2\pi \widetilde{r}_A}$ which resembles the temperature of a spherically symmetric black hole with horizon radius $\widetilde{r}_A$ \cite{dd}.\par
The modified Bekenstein-Hawking entropy in Rastall gravity on the apparent horizon is given as \cite{rastall1, rastall2}
\begin{equation}
\widetilde{S}=\left(1+\frac{2\gamma}{4\gamma-1}\right)S_0 ,
\end{equation}
with $S_0=\frac{A}{4}$ is the Bekenstein-Hawking entropy on the apparent horizon and $A=4\pi \widetilde{r}_A^2$ is the area. The normal Bekenstein-Hawking entropy in GR is recovered for $\gamma \rightarrow 0$. So, the area $A=4\pi \widetilde{r}_A^2$ is related to the modified area $\tilde{A}$ by $\tilde{A}=(1+\frac{2\gamma}{4\gamma-1}) A$ where the units $c=G=\hbar=1$ has been considered. As a function of cosmic time $t$, the entropy should always be an increasing function where the Universe evolves to the equilibrium state of maximum entropy. That means the two conditions
\begin{equation}
\dot{S} \geq 0 ~~~\text{and} ~~~\ddot{S} \leq 0,
\end{equation}
should be satisfied. We recall that for a system at equilibrium, entropy has the maximum value which means that it can't increase anymore. On the other hand, any decrease is not possible as it violates the second law of thermodynamics which states that
\begin{equation}
d(S_m+S_h)>0,
\end{equation}
where $S_h$ is the horizon entropy, and $S_m$ the entropy of the entire matter fields. At the apparent horizon, the FRW equations can be written as $dE = T dS + W dV$  where $E$ is the total energy and $W=\frac{1}{2}(\rho-p)$ is the work density \cite{work1,work2}. The cosmological work density has been related to the cosmological constant in \cite{nascyc}. The thermodynamical quantities $w$, $\widetilde{r}_A$, $T_A=$ and $S$ can be simply expressed as a function of redshift $z$ to probe their behavior. Figure \ref{fn} shows that the work density $W$ has the same physical behavior of energy density $\rho$ vesrses $z$. In terms of cosmic time, it's a decreasing function during expansion which means that the work done by the cosmic volume change due to the change of the apparent horizon radius decreases. The first derivative of the event horizon radius is positive during the expansion $\dot{\widetilde{r}_A}>0$ and negative during the contraction  $\dot{\widetilde{r}_A}<0$. A comparison between the evolution of Hawking temperature $T_H$ and the temperature on the horizon $T_A$ verses $z$ has been plotted in Figure \ref {Fn6}. Entropy $\tilde{S}$ is always positive. While the condition $\dot{\tilde{S}}>0$ is satisfied, the non-positivity of the equilibrium condition $\ddot{\tilde{S}}$ exists only for the second half of cosmic cycle.

\section{Causality and Energy Conditions}

One way to investigate the physical acceptability is testing the classical linear energy conditions (ECs) \cite{hawk,ec12}, as the quantum corrections are ignored in the present model. The null, weak, strong and dominant ECs are respectively: $\rho + p \geq 0$; $\rho \geq 0$, $\rho + p \geq 0$; $\rho + 3p \geq 0$ and $\rho \geq \left|p\right|$. These linear conditions can't be satisfied in the existence of quantum effects \cite{ec}. The strong energy condition (SEC), for example, implies that gravity should always be attractive which is not realistic when describing cosmic acceleration or inflation \cite{ec3,ec4,ec5} . For the present model,
\begin{eqnarray*}
\rho+p&=& -\frac{2}{K}\dot{H} ,   \\   
\rho-p&=&  -\frac{2}{K} (12\lambda K H^2+6\lambda K \dot{H}-3H^2-\dot{H}),    \\
\rho+3p &=&   \frac{6}{K} (4\lambda K H^2+2\lambda K \dot{H}-H^2-\dot{H}) .  
\end{eqnarray*}
The SEC is not expected to be valid during the negative pressure-dominated accelerating era as the negative pressure represents a repulsive gravity effect. The DEC is valid all the time for this model which is expected as it implies the non-negativity of energy density. The SEC is valid only with the domination of attractive gravity in the decelerating era (at the first half of each cycle ), and then becomes invalid with the domination of repulsive gravity during the accelerating era.\par
\begin{figure}[H]
  \centering            
	\subfigure[$W$]{\label{n2}\includegraphics[width=0.3\textwidth]{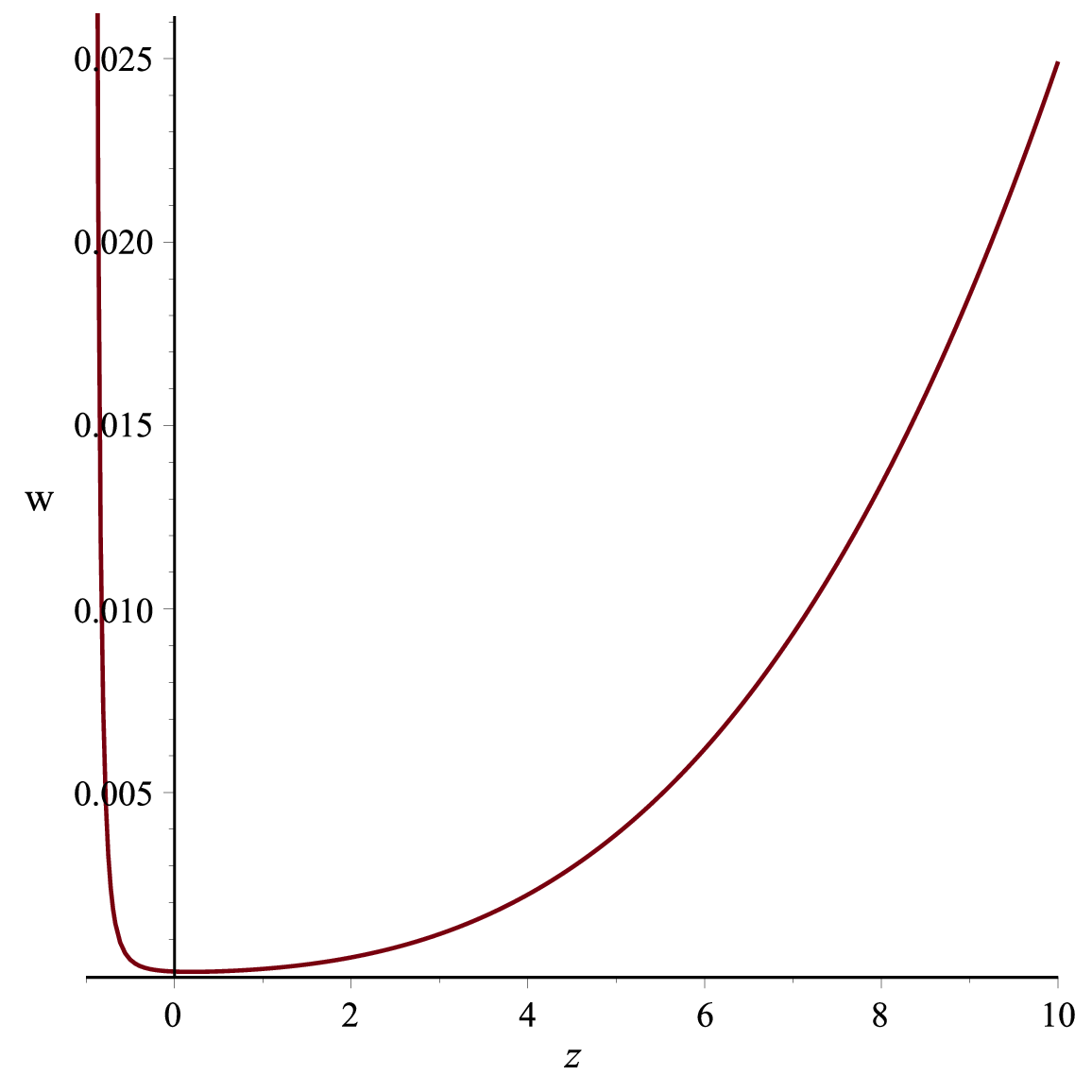}}\hspace{5mm}
		\subfigure[$\widetilde{r}_A$]{\label{n8}\includegraphics[width=0.3\textwidth]{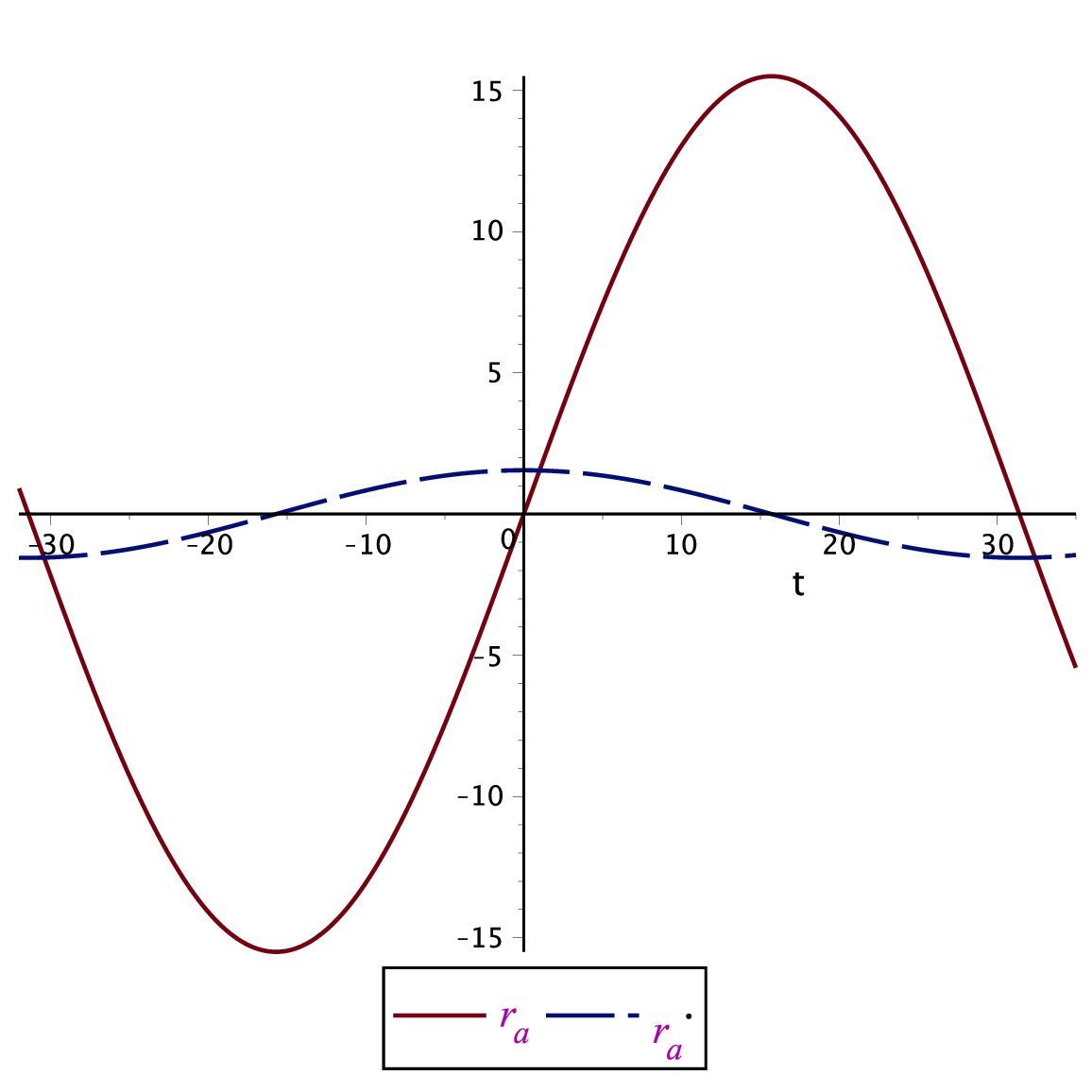}}\hspace{5mm}
	 \subfigure[$T_A$]{\label{Fn6}\includegraphics[width=0.3\textwidth]{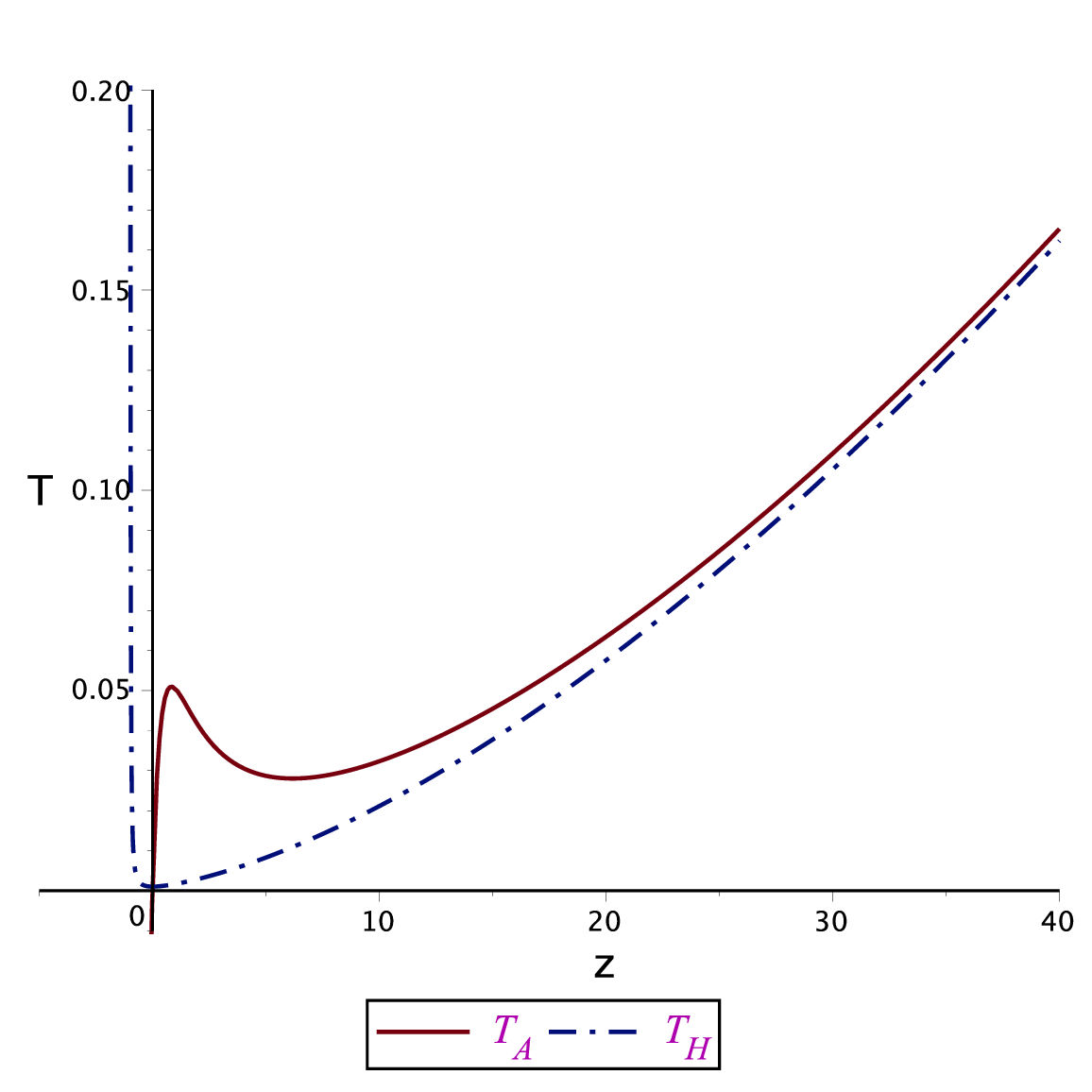}}\\
	 \subfigure[$\tilde{S}(t)$]{\label{F6n}\includegraphics[width=0.3\textwidth]{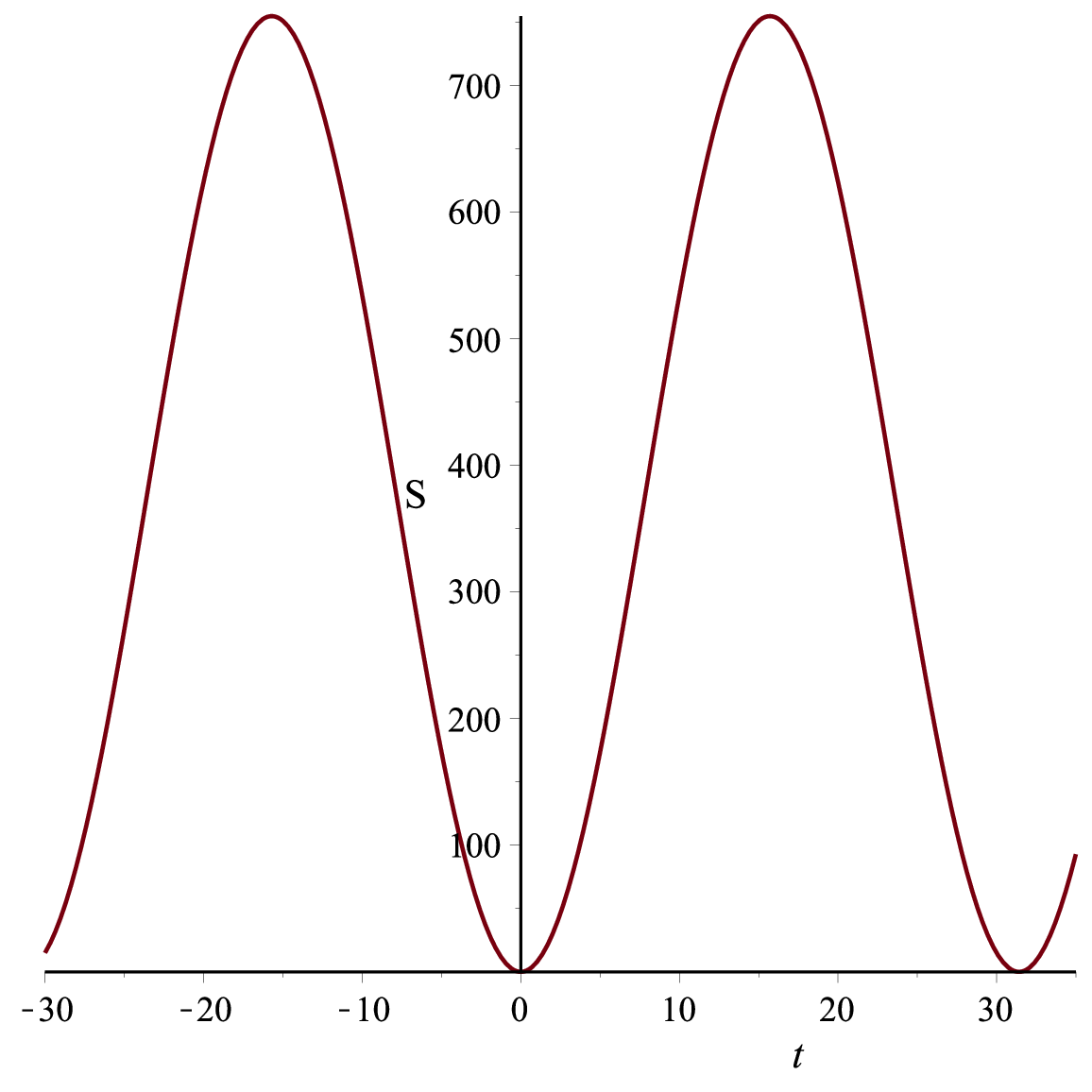}} 	\hspace{5mm}
		 \subfigure[$\tilde{S}(z)$]{\label{n80}\includegraphics[width=0.3\textwidth]{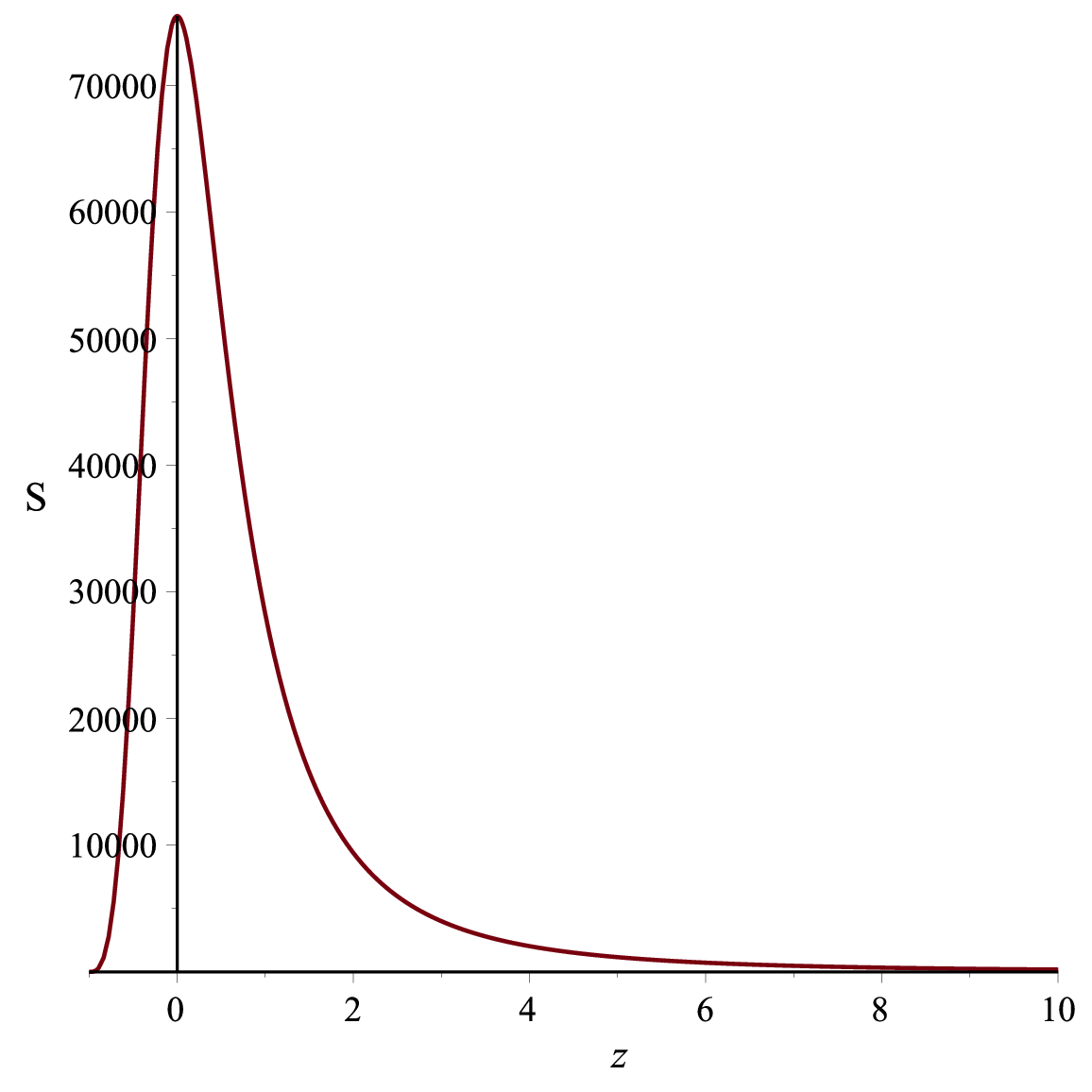}} 	\hspace{5mm}
			 \subfigure[$\dot{\tilde{S}}$ \& $\ddot{\tilde{S}}$]{\label{Fnb}\includegraphics[width=0.3\textwidth]{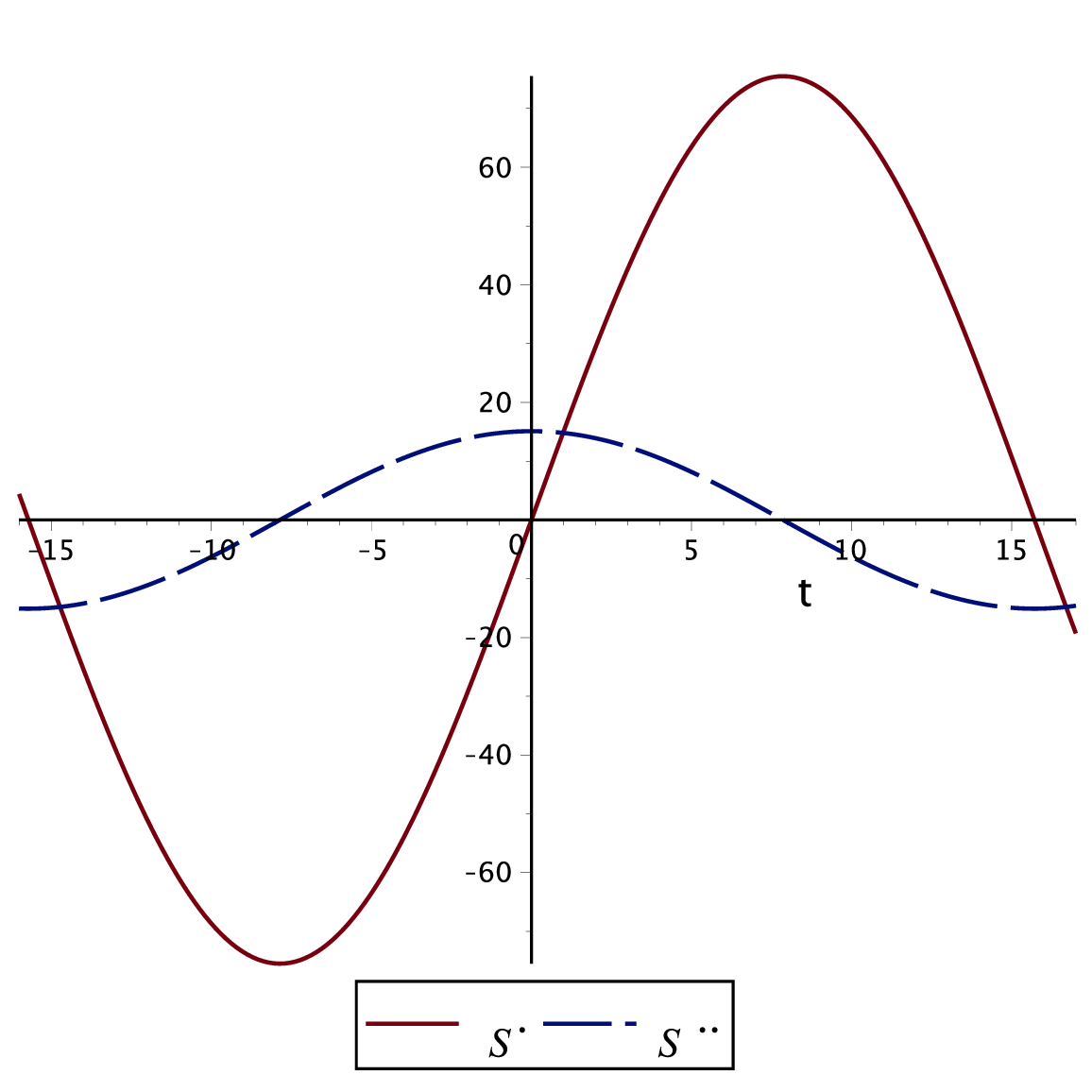}}\\
		 \subfigure[$\dot{\tilde{S}}$ \& $\ddot{\tilde{S}}$]{\label{Fpb}\includegraphics[width=0.3\textwidth]{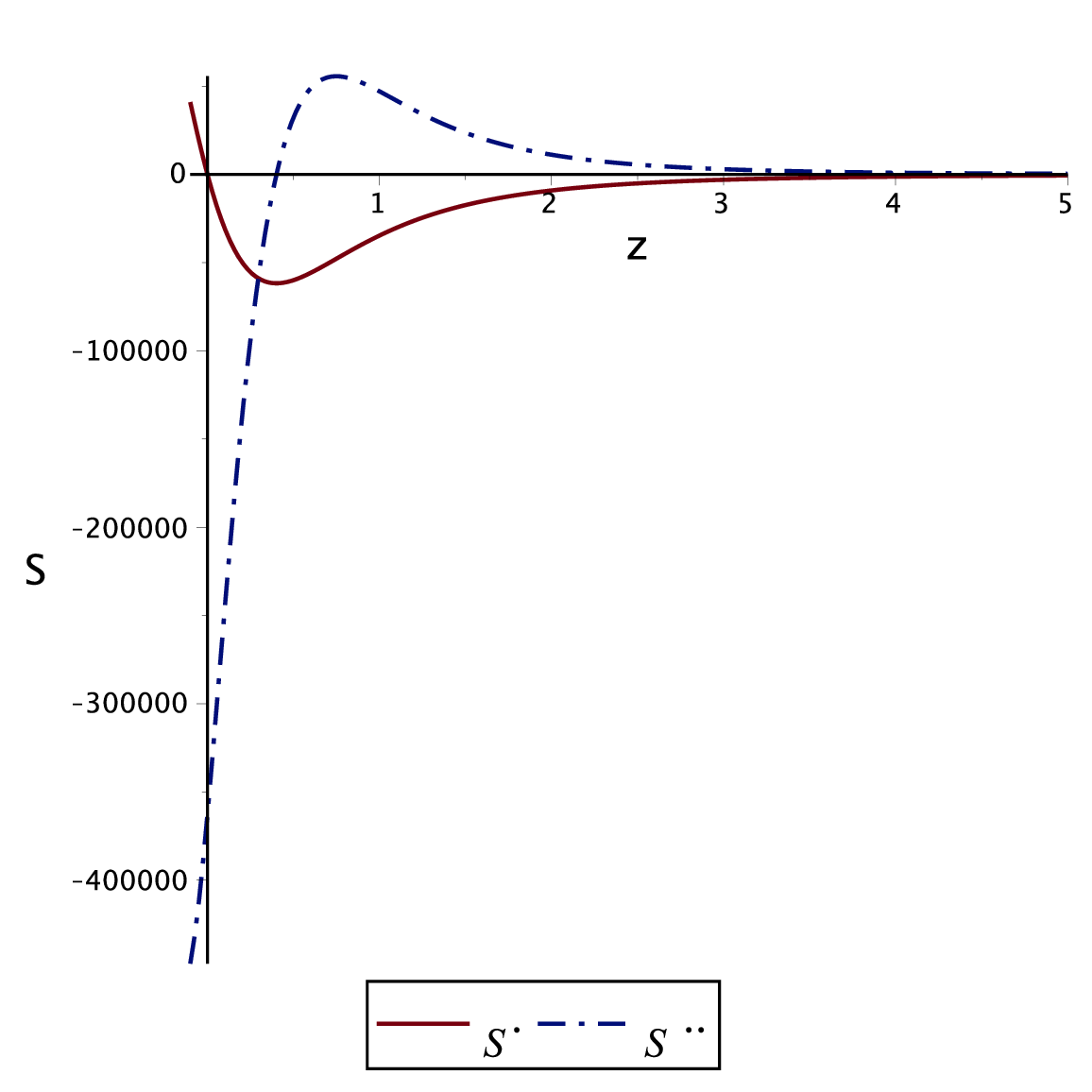}} \hspace{5mm}
	 \subfigure[$C_s^2$]{\label{F0b}\includegraphics[width=0.3\textwidth]{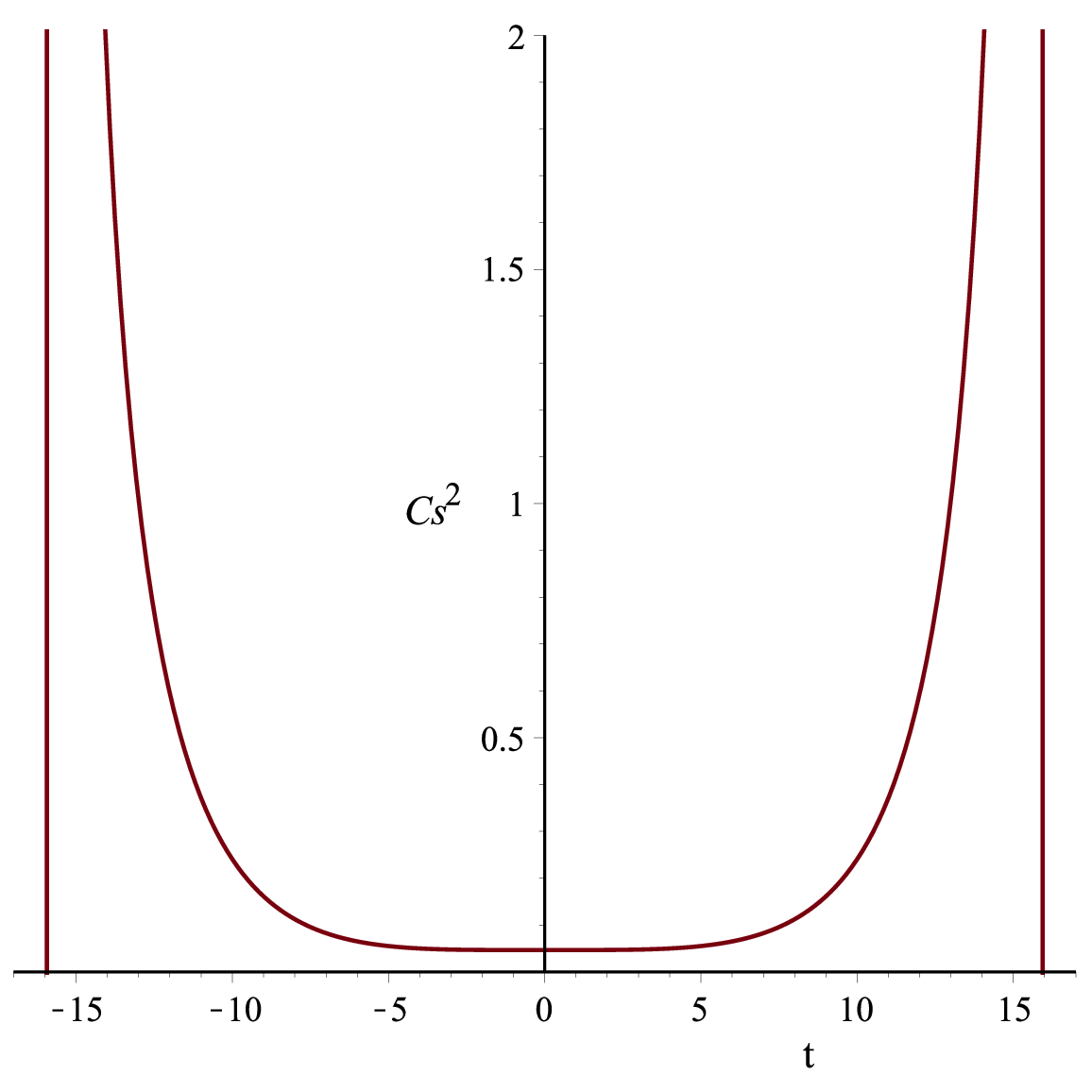}} \hspace{5mm}
	\subfigure[$ECs$]{\label{Fi0b}\includegraphics[width=0.3\textwidth]{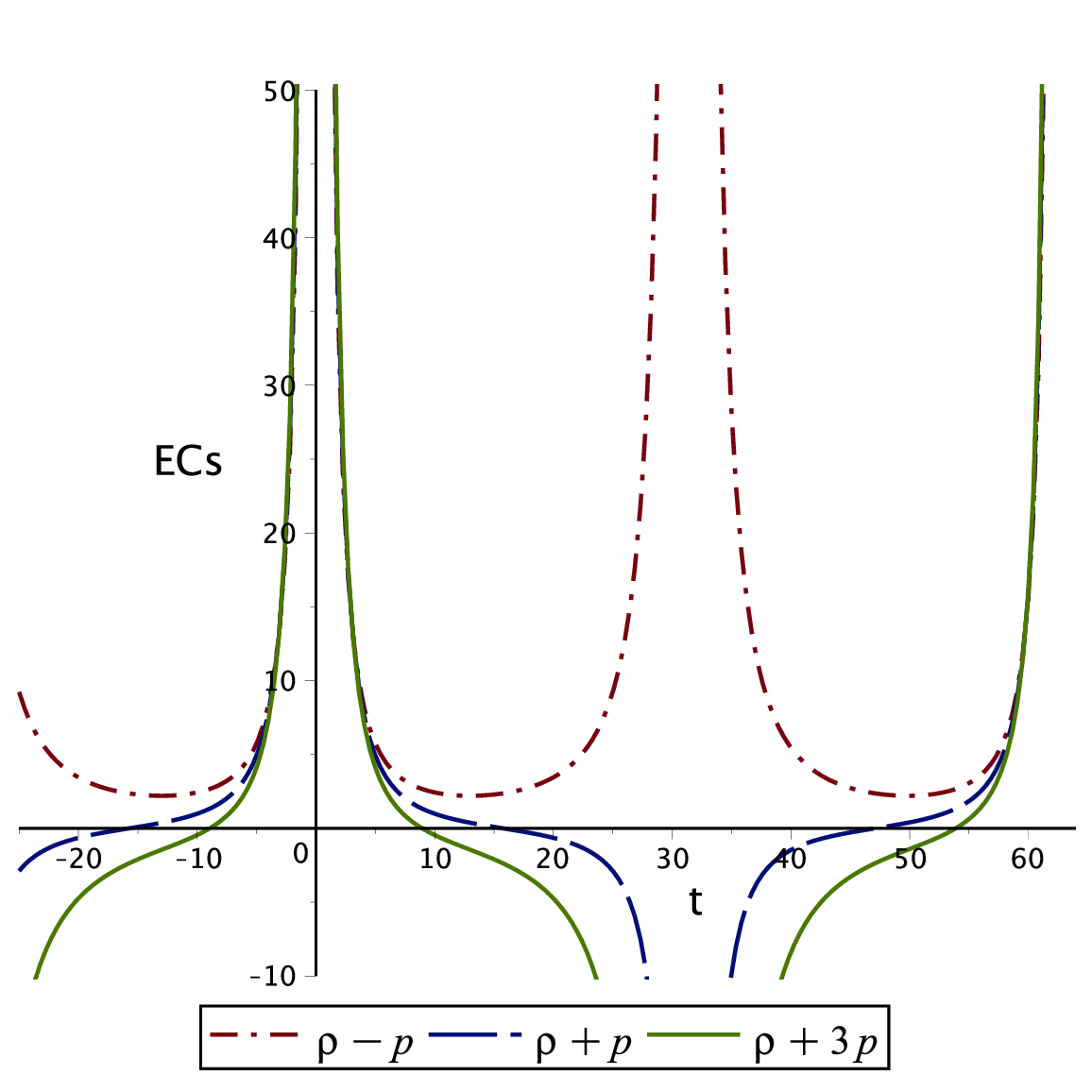}} 
		 \caption{ (a) Work density verses redshift. (b) Radius of the apparent horizon verses redshift. (c) Temperature on dynamical horizon verses Hawking temperature of a spherically symmetric black hole. (d) \& (e) evolution of entropy verses $t$ and $z$ shows that $\tilde{S} \geq 0$. (f) $\dot{\tilde{S(t)}}$ and $\ddot{\tilde{S(t)}}$, while the condition $\dot{\tilde{S}}>0$ is satisfied, the non-positivity of the equilibrium condition $\ddot{\tilde{S}} <0$ exists only for the second half of cosmic cycle. (g) $\dot{\tilde{S(z)}}$ and $\ddot{\tilde{S(z)}}$. A violation of $\ddot{\tilde{S(t)}} <0$ starts at $z \gtrapprox   0.4$ (h) The causality condition $0 \leq \frac{dp}{d\rho} \leq 1$ is satisfied except except near the initial and the future Big Rip singularities (i) Linear energy conditions. Here $m=1.55$, $\lambda=1.4$, $K=0.01$, $k=0.1$ and $a_0=1$. }
  \label{fn}
\end{figure}

The causality condition for the adiabatic square sound speed $0 \leq \frac{dp}{d\rho} \leq 1$  should be satisfied through cosmic evolution. For the current model, using (\ref{9}) and (\ref{10}), we get
\begin{equation}
c_s^2=-\frac{1}{3}\frac{12 \lambda K H \dot{H}+3\lambda K \ddot{H}-3H\dot{H}-\ddot{H}}{4 \lambda K H \dot{H}+\lambda K \ddot{H}-H \dot{H}}.
\end{equation}
Figure \ref{F0b} shows that the causality is satisfied except near the initial and the future Big Rip singularities

\section{Conclusions}
We have presented a cyclic cosmological model of a flat universe with future big rip singularity in the framework of Rastall gravity. The model possesses a quintom beahvior where the EoS parameter crosses the phantom divide line (cosmological constant boundary) at $\omega=-1$. In such oscillatory cosmic evolution, the deceleration-to-acceleration Cosmic transit happens at around $ 8.7~~ \text{Gyr}$. A recent analysis that included data from cosmic chronometers and the Pantheon compilation of supernovae, calculated the transition redshift to be $z_{t} = 0.61^{+0.24}_{-0.16}$ \cite{N1}. Given that the universe is currently 13.8 $\text{Gyr}$ old, this correlates to a transition period of roughly 7.5 to 9.8 $\text{Gyr}$ ago. A transition redshift of $z_{t} = 0.60^{+0.21}_{-0.12}$ was discovered by another study that looked at a parametrization of the deceleration parameter \cite{N2}. This estimate, which places the transition era between 7.7 and 9.6 $\text{Gyr}$ ago, is in good agreement with the previous one. A cosmic deceleration-to-acceleration transition that took place approximately $ 8.7~~ \text{Gyr}$ ago is within the observed range based on these observational estimations. Thus, our conclusion so seems to be in accordance with recent evidence from observations. \\

The EoS parameter $\omega$ meets $-1$ at redshift $z=0$. The model exhibits the correct physical evolution of energy density and a sign flipping of cosmic pressure in compatible with cosmic transit. Observational evidence supports that the EoS parameter $\omega$ is very close to $-1$ at the present epoch (redshift $z= 0$) \cite{N3,N4}.\\

 The first derivative of the Hubble parameter $\dot{H}$ is negative during the expansion which means a decreasing $H$, and positive during contraction which means an increasing $H$. The EoS parameter $\omega$ lies in the range $-2.25 \leq \omega(t) \lesssim \frac{1}{3}$. It begins at the radiation-like era then passes through the dust era, the current dark energy-dominated era before crossing the cosmological constant boundary to the phantom era. In the same time, plotting the $\omega$ evolution in Rastall Gravity (RG) and in General relativity (GR) together shows a tiny difference between the two curves. \\
 
The evolution of the temperature on dynamical horizon and Hawking temperature has been plotted verses redshift. The entropy in the model is positive all the time.  It starts as an increasing function of time in the first expanding half cycle until it reaches maximum value corresponding to an equilibrium state. The rate of change of the entropy along with the equilibrium condition has been calculated and plotted. The causality is satisfied all the time except near the initial singularity and the future Big Rip singularity. \\

In order to solve problems such as the initial singularity and future big rip situations, the oscillating Rastall universe model sets itself apart by including Rastall gravity to produce a cyclic cosmological model with quintom behavior.  However, other research have used alternative processes to help transcend the phantom divide, including parameterized post-Friedmann approaches, two-field models, holographic dark energy models, and gravity modifications like $f(Q)$ gravity.  In order to address the difficulties of simulating the universe's accelerated expansion and the dynamic character of the equation of state for dark energy, each framework provides distinct perspectives and approaches. The Introduction already includes a detailed discussion and the recent references. 
	\\

It is pertinent to investigate an oscillating Rastall universe that crosses the phantom division line because it provides a theoretical foundation for investigating dynamic dark energy models and possible gravity alterations. Therefore, our research is useful and pertinent to the oscillating Rastall cosmos.


\section*{Acknowledgments}
The authors (A. Pradhan \& A. Dixit) are grateful for the resources and support provided during a visit to the Inter-University Centre for Astronomy \& Astrophysics (IUCAA), Pune, India, as part of their Associateship program. In order to improve the manuscript in its current form, the authors would like to thank the anonymous reviewer for his insightful remarks.

\section*{ORCID}
Nasr Ahmed https://orcid.org/0000-0003-1544-9970\\
Anirudh Pradhan https://orcid.org/0000-0002-1932-8431 \\ 
Archana Dixit https://orcid.org/0000-0003-4285-4162  


\end{document}